# Adaptive modulations of martensites

S. Kaufmann[1,2], U.K. Rößler[1], O. Heczko[3,1], M. Wuttig[4], J. Buschbeck[1], L. Schultz[1,2] and S. Fähler[1,2]

[1] IFW Dresden, P.O. Box: 270116, 01171 Dresden, Germany

[2] Institute for Solid State Physics, Department of Physics, Dresden University of Technology, 01062 Dresden, Germany

[3] Institute of Physics, Academy of Science of Czech Republic, Na Slovance 2, 182 02 Prague, Czech Republic

[4] Department of Material Science, University of Maryland, College Park, MD, 20742, USA

Modulated phases occur in numerous functional materials like giant ferroelectrics and magnetic shape memory alloys. To understand the origin of these phases, we review and generalize the concept of adaptive martensite. As a starting point, we investigate the coexistence of austenite, adaptive 14M phase and tetragonal martensite in Ni-Mn-Ga magnetic shape memory alloy epitaxial films. The modulated martensite can be constructed from nanotwinned variants of a tetragonal martensite phase. By combining the concept of adaptive martensite with branching of twin variants, we can explain key features of modulated phases from a microscopic view. This includes phase stability, the sequence of 6M-10M-NM intermartensitic transitions, and magnetocrystalline anisotropy.

In numerous materials, electric and magnetic fields can distort lattice unit cells, resulting in an associated strain which is commonly below 0.3 %. Recently, ferroelectric [1] and magnetic shape memory materials [2] have been found where applied fields control the orientation low symmetry unit cells. Thus giant strains of several percent are achieved by a rearrangement in twinned microstructures. Why relatively weak electrical or magnetic fields can change the microstructure of a solid material is not fully understood. However, in both material classes, these effects take place only in phases with a modulated structure [3,4].

Commonly, such modulated structures are considered as thermodynamically stable phases. The displacive transition from a high-symmetry austenite to a low-symmetry phase is of the martensitic type. The transformation requires that this martensite is accommodated on a habit plane as lattice invariant interface which fixes the geometrical relationship between the two crystal structures [5]. The lattice mismatch is compensated by twinning of the martensite. Hence a large number of twin boundaries, connecting differently aligned martensitic variants are introduced. Extrapolating this geometrical continuum approach to the atomic scale, Khachaturyan et al. [6] argue that the modulated structures observed in materials with lattice instabilities should be understood as ultrafinely twinned metastable structures and not as thermodynamically stable phases. In this view, the large and complex apparent unit cell of the modulated phase is composed of nano-twin lamellae of a simpler, thermodynamically stable, martensitic phase. The twinning periodicity and, hence, the modulation is determined by geometrical constraints and the transformation path. The key requirement for the validity of this explanation is very low nano-twin boundary energy. The power of this concept is demonstrated for the Ni-Mn-Ga system, where the sequence of phase transitions, magneto-crystalline anisotropy and microscopic models for intermartensitic transitions as well as stress induced martensite are explained and generalized.

This concept of adaptive martensite competes with alternative theoretical ideas, e.g., emphasizing the relevance of Fermi surface nesting for modulated phases in metallic martensites [7,8]. Direct experimental proofs for the adaptive nature of modulated martensite structures are difficult because thermodynamic measurements do not easily identify metastable phases. Furthermore, diffraction experiments cannot directly distinguish a regular nano-twinned microstructure from a long-period modulated phase [9]. For lead-based ferroelectric perovskites, the concept of adaptive phases has been employed to explain the transitional region at the morphotropic phase boundary [3,10,11]. Still, the adaptive concept is strongly debated [12]. One approach explains anomalous phenomena at the morphotropic phase boundary by the existence of low symmetry equilibrium phases as bridging structures [13]. This assumption, however, cannot explain the transformation paths between, and the co-existence of these different modulated and non-modulated structures.

In this letter, starting from experimental observations on epitaxial NiMnGa films, we demonstrate that the 14M modulated phase observed in bulk [14] is a metastable adaptive phase. Since this modulated phase exhibits an anomalously large strain due to twin re-arrangement under magnetic fields, adaptivity seems to be crucial for the giant strain effects not only in the ferroelectrics but also in magnetic shape memory materials. We identify epitaxial films as a suitable experimental setting to decide on the origin of modulated phases. We exploit two key advantages of epitaxial films as compared to bulk: First, the geometrical constraint at the interface to the rigid substrate stabilizes otherwise thermodynamically unstable phases rendering frozen intermediate states accessible to experimention. Second, the single crystalline substrate acts as a reference system which allows probing crystallographic orientations of all phases in absolute coordinates.

As a model system to test Khachaturyan's concept we selected the Ni-Mn-Ga magnetic shape memory alloy. For the chosen alloy composition, the modulated 14M lattice cell is built from unit cells of the thermodynamically stable non-modulated (NM) phase. The geometrical martensite theory [5] predicts a periodic twinning of the tetragonal martensite lattice, expressed through the fraction of the twin lamella widths $d_1$ and $d_2$: $d_1/d_2 = (a_{NM}-a_A)/(a_A-c_{NM})$. Here, $a_{NM}$ and $c_{NM}$ represent the lattice constants of the tetragonal martensite and aA the lattice constant of the cubic austenite. This ratio directly determines the minimal number of stacked atomic layers forming the adaptive phase. In diffraction experiments, one expects an orthorhombic lattice with: $a_{14M} = c_{NM} + a_{NM} - a_A$, $b_{14M} = a_A$ and $c_{14M} = a_{NM}$. These relationships can be understood by describing the modulated martensite as a periodic superlattice [9]. Diffraction on such a periodic structure results in satellite reflections appearing like a new phase. The connection between the nanotwinned modulated structure and a macroscopic amount of NM phase can be described by branching [15] which preserves the orientation relationship between the austenite and the NM twins. The adaptive nature of 14M explains why this structure does not correspond to a global energy minimum in first-principles calculations, which consistently find the tetragonal NM phase as ground state [16].

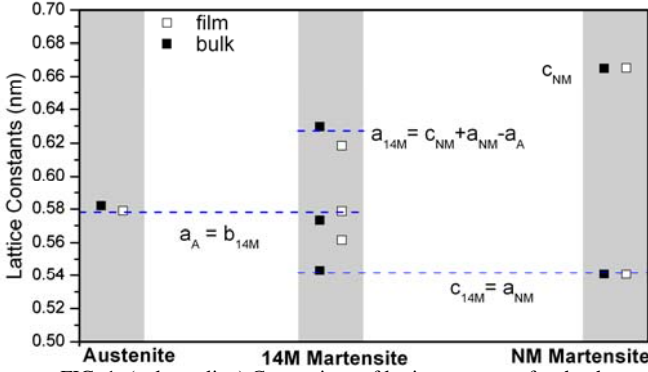

FIG. 1: (color online) Comparison of lattice constants for the three different phases. The blue dashed lines mark the calculated lattice constants of the adaptive martensite phase. For each phase, film and bulk [14] lattice constants are shown.

We have investigated epitaxial Ni-Mn-Ga films with a thickness of about 500 nm, deposited by DC magnetron sputtering on MgO(100) substrates at 250°C. Structural characterization was performed by X-ray diffraction (XRD) in a Philips X'Pert 4-circle setup with Cu-K$_\alpha$ radiation. θ–2θ-scans of the {400}-planes, performed at room temperature [17], reveal the coexistence of three phases: the cubic austenite ($a_A = 0.578$ nm), the pseudo-orthorhombic 14M martensite ($a_{14M} = 0.618$ nm, $b_{14M} = 0.578$ nm, $c_{14M} = 0.562$ nm) and the tetragonal NM martensite ($a_{NM} = 0.542$ nm, $c_{NM} = 0.665$ nm). Lattice constants of all phases are described with reference to the cubic L2$_1$ Heusler unit cell. In equilibrium, the coexistence of three phases at one temperature is contradicting the Gibbs phase rule, confirming that the interface to the rigid substrate is influencing these diffusionless martensitic transformations.

For both, bulk [14] and the present thin film, the measured lattice parameters of the 14M phase agree with the predicted lattice constants from the concept of adaptive martensite (FIG. 1). Although in thin films the tetragonal distortion of the adaptive phase is less than expected from theory, for bulk and thin films $b_{14M}$ is almost identical to $a_A$. In a simplified geometrical model [6], this equality is the key precondition for a coherent austenite-martensite interface. Thus the most important relation between the lattice constants of the adaptive martensite and austenite is fulfilled. Using the measured lattice constants the concept also predicts a twinning periodicity of $d_1/d_2 = 0.428$ for bulk and $d_1/d_2 = 0.417$ for the thin film. This is close to the ideal value of $d_1/d_2 = 2/5 = 0.4$ expected for the 14M phase consisting of twin variants with widths of 2 and 5 atomic layers, respectively. The difference in $d_1/d_2$ between 0.4 for a perfect $(5\bar{2})_2$ stacking and the value calculated from the measured lattice constants suggests that the structure exhibits stacking faults [6]. This could explain the difference between measured and expected lattice constants.

Applying these results on the lattice geometry we can construct the 14M unit cell by using NM unit cells as building blocks. In addition to a projection shown in FIG. 2, a foldable 3D model is available as EPAPS document [17]. We start from NM martensite unit cells, originating from tetragonal deformed and slightly rotated L2$_1$ austenite. One NM unit cell is exemplarily marked with grey background; the different variants are connected by (101)-type twin boundaries. In FIG. 2 one also finds the commonly used 14M unit cell described in the "bct" system [16] (rotated by about 45° to the NM cells and marked with yellow background).

The only difference of our expanded picture compared to the common picture is, that one can directly identify the nanotwinned NM variants. Using the NM lattice constants and the $(5\bar{2})_2$ twinning periodicity one can calculate the lattice constants of the adaptive 14M phase by basic geometry. Additionally the angles between crystal axes of the tetragonal NM twin variants and the axes of the 14M unit cell can be calculated (as sketched in FIG. 1). Lattice constants ($a_{ad}^{bct} = 0.428$ nm, $b_{ad}^{bct} = 2.955$ nm, $c_{ad}^{bct} = 0.542$ nm) and monoclinic angle ($\beta = 95.3°$) in the commonly used bct reference system are close to bulk measurement data ($a_{14M}^{bct} = 0.426$ nm, $b_{14M}^{bct} = 2.954$ nm, $c_{14M}^{bct} = 0.543$ nm, $\beta = 94.3°$) [14].

The identification of a 14M martensite with a nanotwinned NM martensite suggests that macroscopic NM variants are connected to the nanotwinned NM variants by a branching mechanism [15], which does not change the orientation of the NM variants. Thus, the angles between axes of 14M and its NM building blocks should be the same as those between the 14M and the macroscopic NM martensite.

This can be proved using an advantage of epitaxial films where the substrate provides a fixed reference frame. Thereby, it is possible to study the different crystallographic orientations in absolute coordinates by investigating {004} pole figures of all different phases and orientations. These pole figures give the real-space orientation of the different unit cells selected by their lattice spacing. Following our previous report [18] we know that the orientation of the austenite is determined by the epitaxial growth. Moreover some austenite phase remains at the interface to the rigid substrate below the martensitic transformation temperature [19]. For the present film the austenite contributes to the $(400)_A + (040)_{14M}$ pole figure which exhibits one intense peak at zero tilt (FIG. 3 (a)). Since $a_A$ and $b_{14M}$ are equal, as predicted from the adaptive martensite concept, both lattices contribute to the same pole figures.

In contrast to the austenite, the 14M pole figures also exhibit peaks at tilted positions. The orientations of these poles are almost identical to the ones reported previously for an epitaxial 14M film [18]. These orientations agree with the predictions of Wechsler, Liebermann and Read (WLR) theory of an almost exact habit plane [5]. Hence the orientation of the 14M martensite variants is completely determined by the requirement of a coherent interface to the austenite.

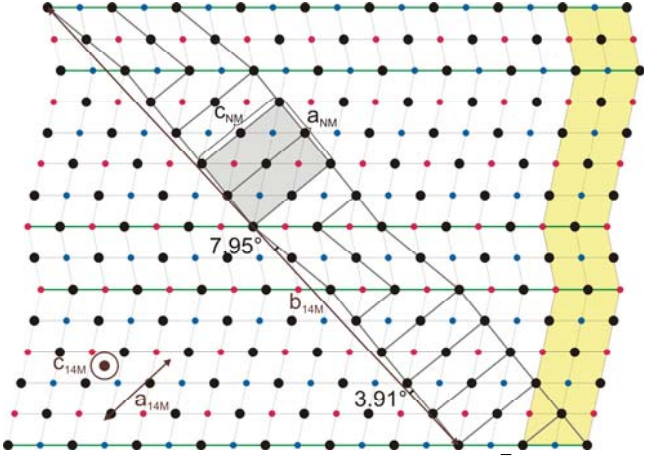

FIG. 2: (color) 14M structure constructed by periodic $(5\bar{2})_2$ twinning of tetragonal NM building blocks. One of the NM cells is exemplarily marked with grey background. The directions of the three different 14M lattice parameters are sketched with brown colour. The angles of the NM unit cells subtended with the 14M supercell (thick lines) are given. The conventional unit cell used to describe 14M within the bct reference system is marked with a yellow background at the right.

Compared to 14M, the pole figures of NM martensite (FIG. 3(d),(e)) exhibit significantly more peaks. This is expected since single NM variants cannot form an exact interface to the austenite. Peak positions are summarized in Table I Using the orientation of 14M and the angles given in FIG. 2 we can directly calculate the orientation of NM variants and thus the peak positions in the pole figures (FIG. 3 (d),(e)). The calculated and measured angles agree within the accuracy of the texture device for $\psi$ of ~1°. This confirms that the orientation of the NM variants does not change during coarsening from nanotwinned to macroscopically twinned NM variants. Each reflection in the pole figures of the NM martensite can be assigned to a well defined variant of the nanotwinned structure, which forms the 14M. Starting at macroscopic NM variants, branching of twin boundaries occurs within the NM phase when the habit plane is approached. Since both phases coexist, branching must continue down to the atomic scale.

These experiments confirm that 14M is an adaptive phase. In the following we will use the concept of building modulated phases from a nanotwinned martensite to analyse several peculiarities of systems exhibiting giant strain. To allow a direct comparison with the present experiments, the focus will be on the Ni-Mn-Ga system.

Density functional calculations for the energy curve of $Ni_2MnGa$ as a function of the tetragonal distortion [16] show a global energy minimum at a $c/a = 1.25$. Compared to this NM martensite, the 14M structure exhibits a higher energy. Using the concept of adaptive martensite, we can interpret this energy difference as twin boundary energy $\gamma$ of the NM phase. On the basis of these calculation [16] we derive a twin boundary energy of about $\gamma = 2$ meV/Å$^2$, close to $\gamma = 0.87$ meV/Å$^2$ recently observed in NiTi nanocrystals [20]. This low value of the twin boundary energy fulfils the key requirement for the formation of adaptive martensite in Ni-Mn-Ga.

These above energy considerations suggest that the 14M phase is not thermodynamically stable. Since, however, bulk single crystals exist and can actuate for several $10^6$ cycles [21], 14M can be considered a metastable phase. Metastability requires the existence of an energy barrier hindering the transition from the nanotwinned to the macroscopically twinned NM martensite. This energy barrier may be related to the repulsive forces between twin boundaries and lattice defects like dislocations required for the annihilation of twin boundaries. Since the nanotwinned 14M can easily adapt internal and external stress by variations of the stacking sequence, appropriate processing (e.g. cooling under load) may be a precondition for the formation of a metastable 14M phase. For the present thin films the constraint by the rigid substrate additionally hinders detwinning and thus explains the coexistence of austenite, 14M and NM martensites. Temperature dependent XRD measurements [17] indeed reveal that all phases

Table I: Comparison between calculated and measured crystal orientations for the tetragonal NM martensite variants. The angles $\psi$ and $\varphi$ are sketched in FIG. 3(a). The grey columns mark the three different underlying 14M martensite variants, from which the two differently oriented macroscopic NM variants originate by coarsening.

|   |   | meas. | from $(400)_{14M}$ | meas. | from $(040)_{14M}$ | meas. | from $(004)_{14M}$ |
|---|---|---|---|---|---|---|---|
| $(400)_{NM}$ | $\psi$ | 7.4° | 8.4° | 3.1° | 3.9° | 3.1° | 2.7° |
|  | $\varphi$ | ±18° | ±26.3° | 45° | 45° | 45° | 45° |
| $(004)_{NM}$ | $\psi$ | 5.3° | 4.7° | 7.6° | 7.9° |  |  |
|  | $\varphi$ | 0° | ±10.4° | 45° | 45° |  |  |

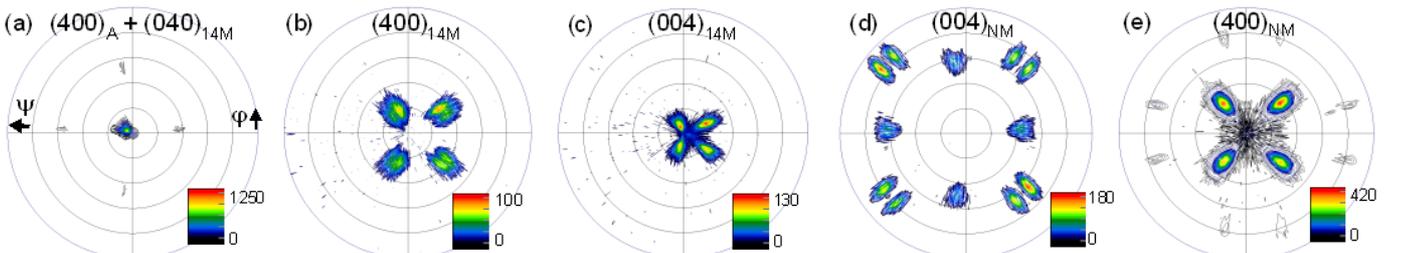

FIG. 3: (color) Pole figure measurements of the {400}-planes of the pseudo-orthorhombic 14M ((a)-(c)) and the tetragonal NM martensite ((d)-(e)) from $\psi = 0\ldots10°$. The four-fold symmetry verifies epitaxial growth on the MgO(100) substrate. The shift of reflections with respect to the centre is due to a slight misalignment of the sample during measurement.

coexist over a temperature range of more than 100 K.

The adaptive phase concept enables to estimate the magnetocrystalline anisotropy of the Ni-Mn-Ga 14M martensite. Since the thickness of the nanotwin variants is significantly below the magnetic exchange length, the magnetocrystalline anisotropy equals the weighted mean of the differently aligned NM variants. Referring to the $L2_1$ system and using data at 300 K [22], the building block concept predicts $K_a^{14M} = -5/7 \cdot K_1^{NM} = 1.63 \cdot 10^5$ J/m$^3$ and $K_b^{14M} = -2/7 \cdot K_1^{NM} = 0.65 \cdot 10^5$ J/m$^3$, using the anisotropy constant $K_1^{NM} = -2.28 \cdot 10^5$ J/m$^3$ of the NM phase. These values agree with the measured constants $K_a^{14M} = 1.72 \cdot 10^5$ J/m$^3$ and $K_b^{14M} = 0.83 \cdot 10^5$ J/m$^3$. This favourable comparison also holds for their temperature dependence [22]. The alignment of the hard axis in 14M is correctly predicted as inclined by 45° with respect to the twinning planes of the NM nanotwins and not by 0° or 90°, as expected by interface anisotropy.

In Ni-Mn-Ga, the phase sequence austenite, 6M (premartensitic), 10M (5-layer), 14M (7-layer), NM martensite is commonly observed with increasing electron density [23]. Detailed diffraction experiments on the modulated structures reveal that they own orthorhombic or monoclinic unit cells [24, 25]. In analogy to 14M all these structures can be assembled from tetragonal building blocks. Their tetragonality increases with the electron density ($c/a_{NM}$ = 1.0152 (for 6M), 1.16 (for 10M), 1.26 (for 14M) [17]. Since typically Ni-Mn-Ga alloys exhibit the same phase sequence during cooling [23], we suggest that the electron density is the key parameter which controls the tetragonal distortion. This parameter is systematically varied by stoichiometry and/or thermal expansion. All the modulated phases can form an almost exact habit plane to the austenite [17]. This indicates that the austenite-martensite interface energy σ and the nano-twin boundary energy γ are small and similar in magnitude for all phases. Hence, the interface energies do not appreciably influence the relative stability of the modulated phases. Therefore, all modulations observed in the Ni-Mn-Ga system are adaptive. The sequence of the modulated structures 6M – 10M – 14M is determined by the electron density e/a via the variation of the tetragonal distortion in the equilibrium NM phase. In the metastable modulated phases, the nano-twin widths are minimized for $(c/a)_{NM}$ values which results in a fraction of small integer numbers for $d_1/d_2$.

Metastability of the adaptive phase can explain irreversibilities of intermartensitic transitions under external stress. As observed in Ni-Mn-Ga alloys, the application of a sufficient external compression selects NM variants with their short axes in compression direction. Under sufficient load nanotwin boundaries vanish, resulting in a transformation to macroscopic NM variants, while a clear reverse transition is not observed [26]. The nanotwinned nature of a modulated martensite can also affect the thermal hysteresis of the martensitic transition. As reported by Cui et al. [27], the mismatch between austenite and martensite lattice constants determines the width of the hysteresis. For an adaptive phase, the precondition of an exact habit plane is fulfilled, thus, a small hysteresis is expected and measured [4].

To conclude, our investigations show that modulated phases in Ni-Mn-Ga originate from the adaption of a thermodynamically stable martensite to the austenite. This establishes magnetic shape-memory alloys as an important metallic counterpart to ferroelectrics near the morphotropic phase boundary. The similarity between these systems suggests that adaptivity is crucial for field-induced giant strains in martensitic functional materials. The modulated phases facilitate adaption to external forces and fields by a redistribution of nano-twin boundaries, in contrast to a thermodynamically stable, stiff martensite.

The authors would like to thank A.N. Bogdanov, P. Entel, M. E. Gruner and A. G. Khachaturyan for discussions. We gratefully acknowledge funding by the DFG via the priority program SPP1239 www.MagneticShape.de.

# EPAS document for: Adaptive modulations of martensites

S. Kaufmann, U.K. Rößler, O. Heczko, M. Wuttig, J. Buschbeck, L. Schultz and S. Fähler

**1) 3-D model of 14M martensite**

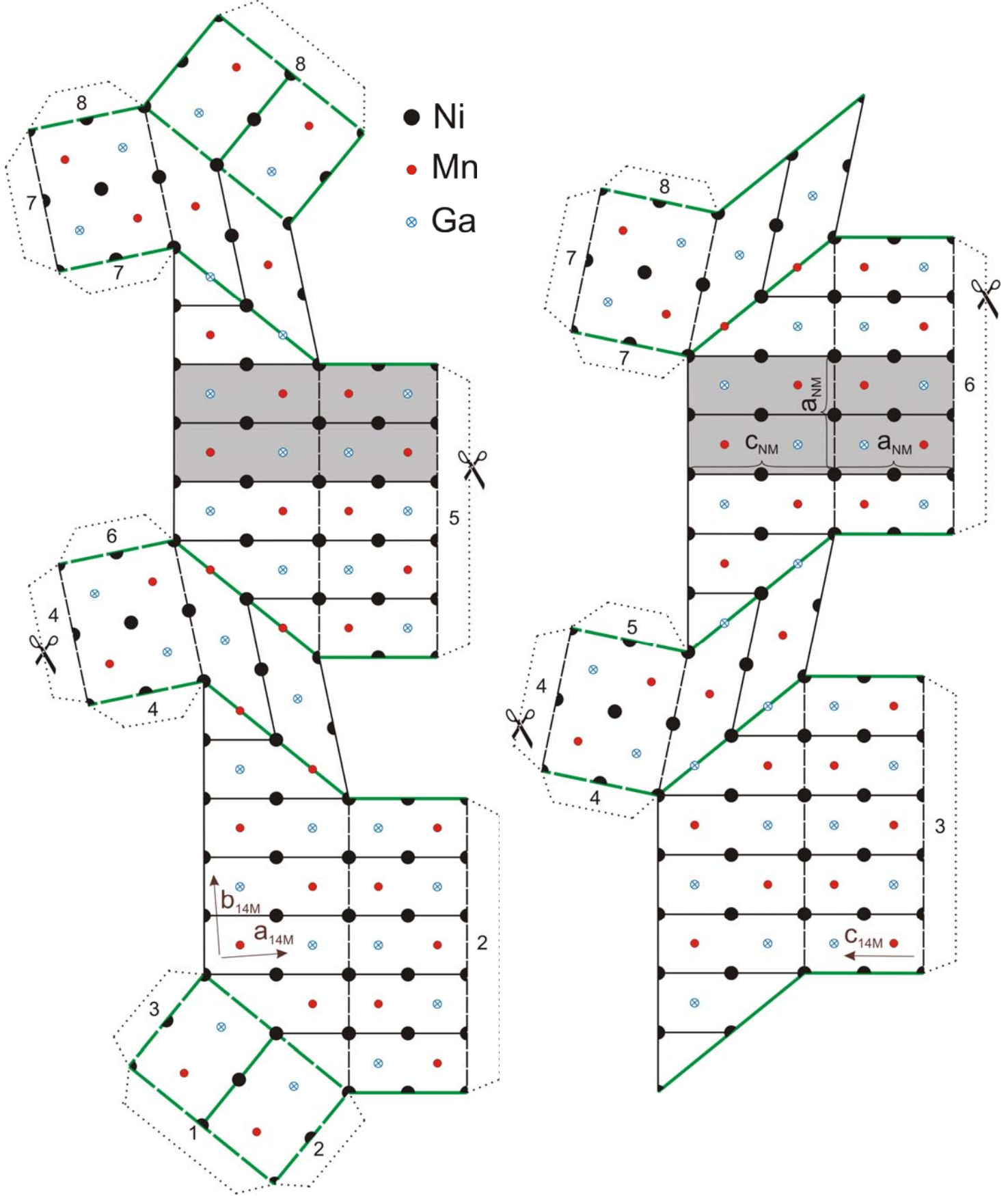

*FIG S1: 3-D model of one 14M supercell of $Ni_2MnGa$, consisting of two different NM variants. For a better understanding of the different tilts determining the orientations of the NM martensite variants, the two counterparts of the figure can be cut out and should be glued following the numbers. Folding edges are marked by dashed lines. One unit cell of the NM martensite is exemplarily marked in grey. Additionally, the directions of the 14M unit cell axis are sketched in brown. The twin boundaries connecting two different NM variants are marked in green. Mn and Ga atoms are not in plane but shifted by ¼ times the NM lattice constant into the interior of the cell.*

The assembled 3D model for the 14M supercell can be used for direct visualization of several key features of the adaptive phase. First however it is helpful to recognize that the complete 14M unit cell is built from NM building blocks. In the five atomic layer thick variant, one NM unit cell has a grey background and its lattice axes are marked. The unit cell is selected in a way, that Ni atoms occupy the edges. Mn and Ga atoms are not in plane but shifted by ¼ of the NM lattice constant into the interior of the cell. The NM unit cell is tetragonal ($c/a_{NM}$ = 1.22). The $(101)_{NM}$-type twin boundaries between the neighbouring NM nanovariants are marked with green lines. The inherent twinning angle results in the characteristic modulated structure. The neighbouring variant is only two layers thick, hence no complete NM unit cell fits into this nanotwin lamella. Therefore it is more convenient to consider half a NM unit cell (framed in black) as building block. Since $Ni_2MnGa$ is an ordered $L2_1$ Heusler alloy, composition of the two twin lamella into $(5\bar{2})$ stacking does not preserve the translation symmetry of the ordered lattice. Therefore, a complete unit cell of the nanotwinned superstructure comprises two $(5\bar{2})$ stacking sequences and has to be described as a 14M modulation. The parameters that fully determine the 14M supercell are chemical order, lattice constants of NM and the $(5\bar{2})$ stacking sequence.

The relevance of this building block principle becomes evident when estimating the magnetocrystalline anisotropy of the 14M martensite. The NM martensite has easy plane anisotropy. The easy plane is spanned by the two $a_{NM}$ axes, while $c_{NM}$ is the hard axis. The magnetocrystalline anisotropy of 14M can now be derived from anisotropy of NM by counting NM building blocks with the hard $c_{NM}$-axis parallel to each crystallographic direction (along $a_{14M}$, $b_{14M}$ and $c_{14M}$) and dividing by the overall number of building blocks. The favoured magnetisation axis of 14M is in $c_{14M}$ direction since no hard $c_{NM}$ axis is aligned in parallel. Parallel to the $b_{14M}$ and $a_{14M}$ directions, fractions of 2/7 and 5/7 of the hard $c_{NM}$ axis are aligned, respectively. Consequently $a_{14M}$ is the hard axis and $b_{14M}$ is semi-hard. With these weighted mean values, one can not only derive the right order of hard, semi-hard and easy axis of 14M, but also the magnitude of the magnetocrystalline anisotropy energies, which agree well with experiments (values are given in the main paper).

**2) Film composition and Structure Analysis**

Composition was determined to be $Ni_{54.8}Mn_{22.0}Ga_{23.1}$ by EDX using a stoichiometric standard. Commonly, bulk samples with similar composition are in NM martensite phase at room temperature [1]. Hence, in this film on MgO(100), austenite and 14M martensite are stabilized by the interface to the rigid substrate.

The coexistence of austenite, 14M and NM martensite was confirmed by XRD θ-2θ measurements. Despite epitaxial growth of austenite at elevated temperature, the martensitic transition results in certain tilts of the martensitic unit cells (See fig. 3). For measuring the lattice constants, in analogy to a previous report [2], sample alignment was optimized for each variant to obtain maximum intensity. In Fig. S1 all five independent measurements are shown in one graph. From these, the lattice parameters given in the main paper were determined. Due to equality of the lattice constants $a_A$ and $b_{14M}$ (expected from the theory of adaptive martensite), the respective peaks coincide.

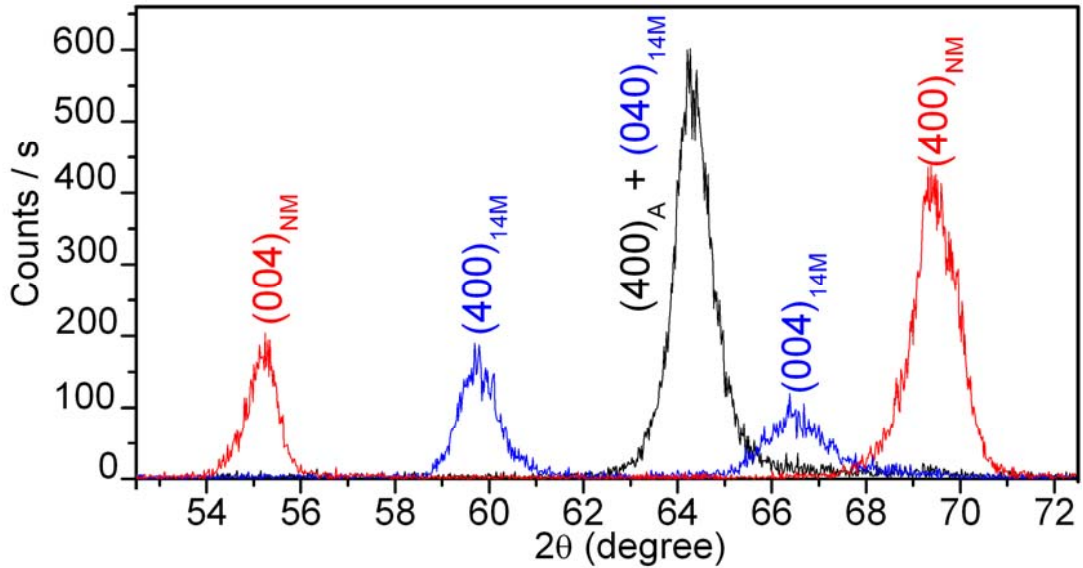

*FIG S2: Summary of X-Ray diffraction patterns (Cu-$K_\alpha$) measured for the different {400}-planes of martensite and austenite, respectively.*

## 3) Temperature dependence

The differences of an adaptive phase compared to a common thermodynamic phase are apparent in temperature dependent X-ray analysis. A qualitative idea about the phase content can be obtained from integrated intensities of reflections of the three different phases. The temperature dependence of the integrated intensities of the {220} reflections for all three phases are displayed in *FIG S3* for the temperature range from 20 to 140°C. In this broad temperature range, all three phases coexist. This observation is in contrast to the properties in bulk systems, where sometimes a well defined sequence of first order (inter)martensitic transitions from austenite to 14M to NM is observed with decreasing temperature [3]. For the present film, the intensity of the reflection from austenite increases continuously with rising temperature. This is expected when approaching the austenite to martensite transformation temperature, however in the accessible temperature range, this transformation is not complete. For the NM martensite the intensity is highest at low temperatures, which is expected for NM being the ground state. The intensity of the 14M martensite shows an unexpected behaviour as it first decreases and then re-increases with rising temperature. For a usual intermediate thermodynamic phase, one would expect phase content and, therefore, maximum intensity near the midpoint of the existence range. The anomalous observation of a minimum intensity at an intermediate temperature indicates that, for the present thin film sample, 14M is not a thermodynamically stable phase, but a unstable, adaptive phase which is sandwiched between austenite and NM martensite.

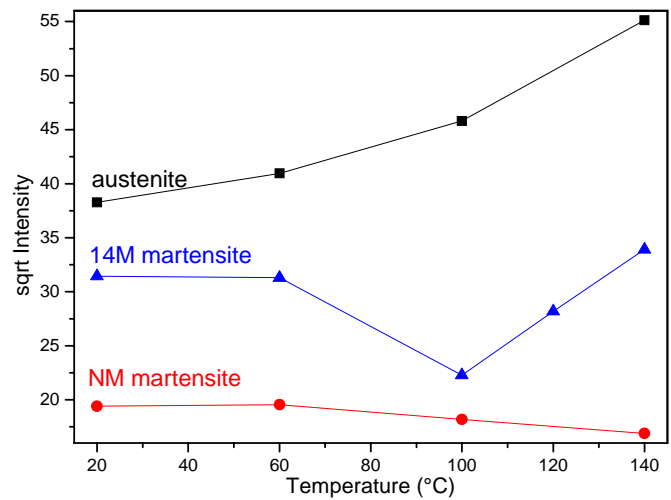

*FIG S3: Transformation of the different phases in the temperature range between 20°C and 140°C. The temperature dependent intensity of the integrated X-ray reflections of the {220}-planes of austenite, 14M and NM martensite, is shown for each phase. The measurements were performed during heating.*

**4) Model of film architecture and calculation of orientation relationship between different phases**

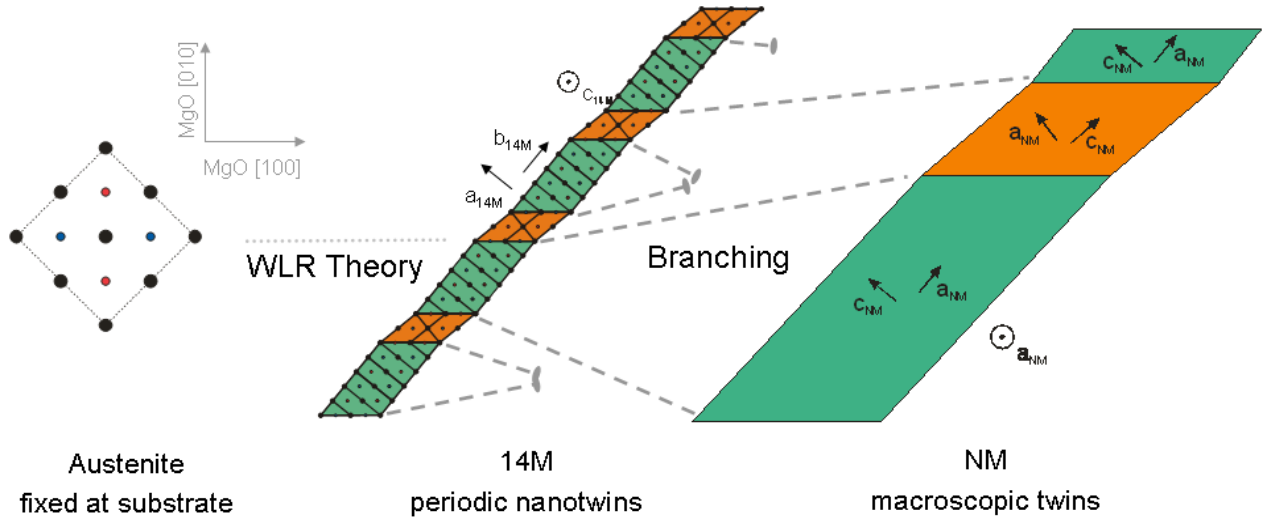

*FIG S4: Sketch of the orientation relationship between Austenite, adaptive 14M and NM phase in constrained epitaxial films. The orientation of Austenite is fixed by epitaxial growth on the MgO(100) substrate (the edges of the MgO unit cell are parallel to the axis system to the left). The geometrical Wechsler-Liebermann-Read (WLR) theory determines the habit plane between Austenite and 14M and hence the orientation of the 14M unit cell. As the nanotwinned 14M structure and the macroscopically twinned NM plate on the same habit plane are distinguished only by the number density of twin boundaries, the orientations of the nano- and macroscopic twin variants are identical. The orientations of the NM crystal axes with respect to the 14M supercell are obtainable by basic geometry, as sketched in Fig. 2 of the main paper.*

In the following we present an approach to describe the film architecture step by step (*FIG S4*). Based on this, the crystallographic orientations between the three observed phases are calculated. In the pole figures, these orientations are quantitatively represented by the peak positions (expressed through the angles $\psi$ and $\varphi$). Due to epitaxial growth, the 14M pole figures exhibit 4-fold symmetry, thus it is sufficient to discuss one of the four equivalent reflections.

From our previous experiments [5] we know that films grow within the austenite state on heated substrates and their orientation is fixed by the epitaxial relationship (MgO(100)[001] || Ni-Mn-Ga (100)[011]). The substrate-film interface hinders the martensite transformation since the constraint at a rigid interface does not allow a variation of the lattice constants. Hence, when cooling below the martensite transformation temperature, some austenite remains close to the substrate [5]. This agrees with the measured pole figure for $(400)_A$ plane, which exhibits one intense peak at zero tilt (Fig 3 (c) in main article). (The additional, weak peripheral reflections in this pole figure are due to second order twinning of $a_{14M}$ and $b_{14M}$ variants).

It is expected that only in direct proximity to the substrate interface, the martensitic transformation is suppressed. Since the film is about 0.5 μm thick, most of the film volume transforms to the martensite state (Fig. S2). For a coherent martensite-austenite interface, the theory of Wechsler, Liebermann and Read (WLR), which is based on the assumption of an invariant plane (the habit plane), can be used to calculate the relative orientation of austenite and martensite [4]. Since the austenite is fixed by the substrate, our pole figure measurements give the orientation relationships in an absolute manner. In a previous work on an epitaxial 14M film we used the measured lattice constants in orthorhombic approximation to calculate the orientation of 14M [5]. These calculated orientations were confirmed by pole figure measurements. The 14M phase of the present film exhibits almost identical lattice constants and orientations (Fig. 4 (a),(b),(c) in the main article). We can conclude that the orientation of 14M is determined by the invariant plane to the austenite at the interface, described by WLR theory. Since an almost exact habit plane is formed, the 14M unit cell is tilted only by small $\psi$ and $\varphi$ angles.

To obtain the orientation of the macroscopic NM variants, we will show in the following that it is sufficient to consider the nanotwinned NM variants forming the 14M supercell. Since the twin boundary angle $\alpha = 90° - 2 \arctan(a_{NM}/c_{NM}) = 11.8°$ between NM variants is fixed by the lattice constants and not by the variant length, annihilation of twin boundaries does not change their orientation. Therefore, the orientation of the NM nanotwin variants forming 14M and those of the macroscopic NM variants are identical, independent of the actual lengths scale of the twin lamellae. As sketched in Fig. 4S, nanotwinned 14M and macroscopically twinned NM can be connected by a branching mechanism [6]. The density of twin boundaries can be reduced successively. This branching mechanism preserves the invariant habit plane on the macroscopic scale. Branching does not leave any degree of freedom for the orientation of the macroscopic NM variants – the orientations of all NM variants are already determined by WLR theory.

The crystallographic orientations of macroscopic NM variants are characterized by two characteristic tilt operations. The first tilt is induced by the orientation of the nanotwinned adaptive lattice with respect to the austenite and described by WLR theory as outlined above. The second tilt is determined by the tilt angle between the crystallographic axes of the nanotwinned NM variants and the 14M supercell axes (as sketched in Fig. 2 of the main paper). Since each 14M variant is built from two NM nanotwin variants, two different sets of peaks are observed in the NM pole figures, whereas only one is observed for 14M. All different combinations of both tilt operations result in the expected reflection positions in the (400)$_{NM}$ and (004)$_{NM}$ pole figures, respectively (Fig. 4 (d),(e) in the main article).

To substantiate this general idea, one can use the 3D model in order to visualize the crystallographic orientations leading to the peaks in the different pole figures of 14M and NM. For this, it is helpful to consider the desk as substrate and place the model with the $c_{14M}$ axis perpendicular to this "substrate plane" and the $a_{14M}$ and equivalently the $b_{14M}$ axes directions rotated by 45° with respect to the desk edges (e.g. as sketched for 14M in *FIG S4*). In order to link the crystallographic orientations with the peak position in the pole figures, please note that in Fig. 4, $\psi$ is only varied by 10°, hence these measurements always reflect the orientation of the axis aligned approximately perpendicular to the substrate. One can illustrate the two different tilt mechanisms as follows: The tilt angles of the 14M variants are always realized by a tilt of the complete 14M supercell around specific axes of the 14M martensite. This tilt mechanism is sufficient to describe the 14M martensite pole figures (FIG. 3 (a-c) in main article). To understand the peak positions in the NM pole figures, additionally, one has to consider the angles between the $b_{14M}$ axis and the inherent NM unit cells. These angles are different for the 5 layer thick (+3.91°) and the 2 layer thick (-7.95°) variant and result from the building block model by basic geometry.

As an example, we will describe the orientation of one particular variant of the 14M martensite and the expected peak positions in the pole figures for the two different NM variants it is connected with. We select the 14M martensite variant with $a_{14M}$ pointing out-of-plane, the orientation of this variants is imaged in the (400)$_{14M}$ pole figure (FIG. 3 (b) in main article). The model has to be rotated by 45° and placed on the desk with $a_{14M}$ pointing out-of-plane. To obtain the tilt towards the substrate normal observed in the (400)$_{14M}$ pole figure, the sample should be slightly tilted around $b_{14M}$ (more precisely by a tilt angle ~2.7° around [010]$_{14M}$). The origin of this tilt is the interface to the austenite as discussed above. By this simple tilt operation, the pole figure of the (400)$_{14M}$ plane is completely described.

The situation becomes more complex when now looking at the macroscopic NM variants, which originated from this 14M variant by branching. Since, as described above, branching does not change the variant orientations, it is sufficient to use the present 3D model. The only difference is that we now consider the orientations of both NM unit cells within the tilted 14M supercell, since both inherent NM variants contribute to different pole figures. We start with the 5 layer thick variant, for which the $c_{NM}$ axis is pointing out-of-plane. Hence this variant contributes to the (004)$_{NM}$ pole figure only. The operations determining the orientation of this NM variant are:

(1) Tilt of the 14M supercell (~ 2.7° around [010]$_{14M}$)
(2) Tilt of the 5-layered variant within the 14M supercell (= 3.91° tilt around [001]$_{14M}$)

Since both tilt operations can also be described in the frame of the 14M unit cell, both tilt axes are perpendicular to each other. Applying both tilt operations successively, one obtains an expected peak position at $\psi$ = 4.8° and $\varphi$ = ±10.4° in the (004)$_{NM}$ pole figure (see also Table 1 in main article). Considering the 4-fold symmetry, the four measured peak positions at $\psi$ = 5.3° and $\varphi$ ~0° in the (004)$_{NM}$ pole figure are explained. The argumentation for the 2-layer thick variant is very similar. As in this case $a_{NM}$ is pointing out-of-plane, this variant contributes to the (400)$_{NM}$ pole figure. The tilt of the 14M supercell is the same as described before, only the tilt of the 2 layer NM variants within the 14M supercell is different (= 7.95° around [001]$_{14M}$). If one applies again both tilt operations, one expects a peak at $\psi$ = 8.4° and $\varphi$ = ±26.3° in the (400)$_{NM}$ pole figure. This corresponds to the peaks measured at $\psi$ = 7.4° and $\varphi$ = ±18°.

The absolute values of the tilt angles can be calculated using the rotation matrix around the unit vector $\vec{v} = \begin{pmatrix} v_1 & v_2 & v_3 \end{pmatrix}^T$:

$$\tilde{D}_{\vec{v}}(\alpha) = \begin{pmatrix} \cos\alpha + v_1^2(1-\cos\alpha) & v_1 v_2(1-\cos\alpha) - v_3 \sin\alpha & v_1 v_3(1-\cos\alpha) + v_2 \sin\alpha \\ v_1 v_2(1-\cos\alpha) + v_3 \sin\alpha & \cos\alpha + v_2^2(1-\cos\alpha) & v_2 v_3(1-\cos\alpha) - v_1 \sin\alpha \\ v_1 v_3(1-\cos\alpha) - v_2 \sin\alpha & v_2 v_3(1-\cos\alpha) + v_1 \sin\alpha & \cos\alpha + v_3^2(1-\cos\alpha) \end{pmatrix}$$

For a direct comparison with the angles $\psi$ and $\varphi$ measured in the pole figure, the substrate edges are used as reference system ($x \triangleq$ MgO[100], $y \triangleq$ MgO[010], $z \triangleq$ MgO[001]). As rotation axis for the first tilt of the complete 14M supercell, one has to use $\vec{v}_1 = \frac{1}{\sqrt{2}}\begin{pmatrix} 1 & -1 & 0 \end{pmatrix}^T$. This vector is identical to the [010]$_{14M}$-axis described in the paragraph above using the Ni-Mn-Ga austenite unit cell as reference system. With $\vec{v}_1$, one obtains $\tilde{D}_{\vec{v}_1}(\alpha)$ with $\alpha$ = 2.7° from WLR theory. This consideration for the first tilt is valid for two of the three 14M variants. The second tilt operation is different for the 2- and 5-layer thick NM variants. The following detailed calculation is carried out for the 5-layer thick NM variant contributing to the (004)$_{NM}$ pole figure (same example as in the paragraph above). First, we have to determine the rotation axis for the second tilt operation. This [001]$_{14M}$-axis depends on the first tilt and can thus be described by the vector $\vec{v}_2 = \frac{1}{\sqrt{2}} \tilde{D}_{\vec{v}_1}(\alpha) \begin{pmatrix} 1 & 1 & 0 \end{pmatrix}^T$ in the MgO reference system. With this rotation axis, one obtains the second matrix $\tilde{D}_{\vec{v}_2}(\alpha,\beta)$, with β =+ 3.9° being the angle between 14M crystal axes and the axes of the inherent 5-layer thick NM variant. The final orientation of the NM martensite variant (described by the vector $\vec{w}$) can be calculated by applying both tilt operations on a vector $\vec{g} = \begin{pmatrix} 0 & 0 & 1 \end{pmatrix}^T$ pointing out-of-plane (MgO[001] direction). Thus, one obtains:

$$\vec{w} = \tilde{D}_{\vec{v}_2}(\alpha,\beta)\, \tilde{D}_{\vec{v}_1}(\alpha)\, \vec{g}.$$

As last step, $\vec{w}$ is transformed from the Cartesian substrate coordinate system to spherical coordinates. The two angle coordinates of $\vec{w}$ directly give the expected peak position ($\psi$ and $\varphi$) in the pole figure. For the 5-layer thick NM variant connected with the (400)14M variant, one obtains $\psi = 4.7°$ and $\varphi = 280.4°$. Considering the 4-fold symmetry given by the substrate, each calculated peak position results in 4 peaks in the respective pole figure. In order to reduce the experimental errors originating from slight sample misalignment, values for the measured angles are averaged over all 4 quadrants. In an analogous manner, we can determine the orientations of the other two 14M variants and the orientations of the connected macroscopic NM variants. The results are summarized in Table 1 in the main article.

## 5) Constructing modulated phases by tetragonal building blocks

At the first glance, this approach already seems to fail when considering 10M martensite since there is no way to construct 10M by using the experimentally observed lattice parameters for the NM phase. As sketched in the following, this however is possible when assuming that NM phases with different $c/a$ ratio may exist. In *FIG S5*, measured and calculated lattice parameters are summarized. As solid horizontal lines the orthorhombic lattice constants of 10M as measured by Righi et al are given [7]. The $c$-axis lattice constant of a virtual NM unit cell is varied whereas its $a$-axis lattice constant is kept constant at $a_{NM} = c_{10M}$, using the direct adaption of the geometrical model for 14M. Assuming a ($3\bar{2}$)-stacking periodicity, the diagonal lines for $a_{10M}$ and $b_{10M}$ are the expected lattice constants for the 10M structure built from the virtual NM. The best agreement of calculation and measurement is obtained at about $c_{NM} = 0.642$ nm. The additionally calculated volume difference of austenite and martensite vanishes for this lattice constant (right y-axis). Hence it seems to be plausible to build 10M from NM building blocks with a $c/a_{NM}$ ratio of 1.161, a significantly lower value than for reported NM martensites, even compared to the large scatter experimentally observed for different NM samples [8].

Additionally, the same idea can be applied on the reported premartensitic phase, which exhibits a 3-layered structure and should thus be called 6M. Investigations by high energy synchrotron radiation revealed an orthorhombic structure with an only very slightly distorted unit cell [9]. Following the approach sketched above, we can calculate the lattice constants of the tetragonal NM unit cells, which can be used as building blocks for the 6M. Comparing the calculated lattice constants with the measured orthorhombic structure, we can determine the lattice constants of the NM building blocks to be $a_{NM} = 0.579$ nm and $c_{NM} = 0.586$ nm leading to a $c/a_{NM}$ ratio of *1.0152*.

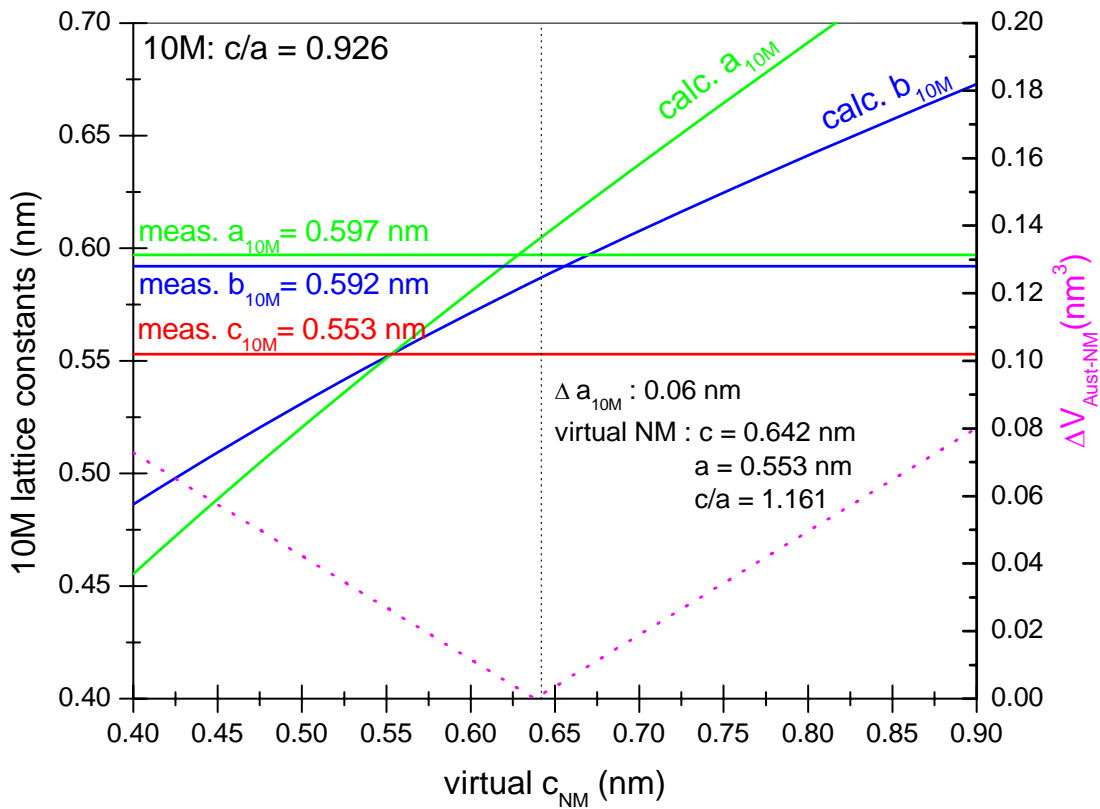

*FIG S5: Approach to construct 10M from NM building blocks with different $c/a_{NM}$ ratio.*

With increasing electron density (e/a ratio) typically the following sequence of martensite phases is observed: A – 6M – 10M – 14M – NM [10]. This indicates that the underlying mechanism for this specific phase sequence is associated with an increase of $c/a_{NM}$ with electron density.

# 6) Martensite-austenite interfaces of the modulated structures in Ni–Mn-Ga

In this section, the interface conditions between the different, modulated phases in the Ni-Mn-Ga system with respect to the cubic austenite are considered. It is shown that all phases allow for the formation of a compatible exact interface, thus, fulfilling the geometrical key requirement to identify these phases as adaptive metastable structures.

This compatibility condition is mathematically expressed by an equation to be solved [11,12],

$$\mathbf{Q}_i \mathbf{U}_i - \mathbf{I} = \mathbf{b} \otimes \hat{\mathbf{m}}, \qquad (0.1)$$

where $\mathbf{U}_i$ describes the deformation from the austenite to one variant of the martensite crystal structure. In Eq. (1.1) the unknowns are the rotation matrix $\mathbf{Q}_i$ and the vectors $\mathbf{b}$ and $\hat{\mathbf{m}}$. These vectors are the so-called shape-strain and the unit normal to the habit plane, respectively. A necessary and sufficient condition that (1.1) has a solution is that the symmetric matrix $\mathbf{U}_i^2$ has an eigenvalue exactly equal one, $\lambda_2 = 1$, one eigenvalue smaller than one, and one eigenvalue larger than one, $\lambda_1 < 1$ and $\lambda_3 > 1$.

In the following we analyse crystallographic structure data for the 6M (3layer premartensite) [9], 10M (5 layer) [7,8], and 14M (7 layer) [8,13] within this formalism. For convenience, we here use the setting for the transformation from a cubic B2-like unit cell into a monoclinically distorted bct unit cell, as in Refs. [12,14]. In Table ST1, we list the lattice parameters of the monoclinic unit cell of the martensite $a_{XM}^{bct}$, $b_{XM}^{bct}$, $c_{XM}^{bct}$, and the monoclinic angle $\beta_{XM}$ for the modulated phases with $X$=6, 10, and 14 and the corresponding lattice parameter of the L2$_1$ cubic lattice cell $a_C$. Then, the resulting eigenvalues, $\lambda_i$, $i = 1, 2, 3$, the vector components of the corresponding habit plane normal $\hat{\mathbf{m}} = (m_1, m_2, m_3)$ and the shape strain $\mathbf{b} = (b_1, b_2, b_3)$ are given that correspond to the variant $\mathbf{U}_1$ in the notation of Ref. [12]. The other habit planes can be obtained by permutation of vector components as detailed in Ref. [12]. From the data in *Table ST1* it is obvious that the parameters of the fundamental bct-like lattice cells are very similar in all modulated phases, a fact already noticed by Pons et al. [8]. Thus, all modulated structures can form a compatible, i.e., almost exact habit plane with the austenite with rather similar orientation $\hat{\mathbf{m}}$ and shape-strain $\mathbf{b}$.

*Table ST1: Compatibility of the austenite-modulated martensite interface for different modulated phases in the Ni-Mn-Ga system according to Eq. (1.1).*

| XM: | 6M | 10M | 14M bulk | 14M epitaxial film |
|---|---|---|---|---|
| Reference | [9] | [7] | [13] | this investigation |
| $a_C$ [nm] | 0.5828 | 0.5825 | 0.5825 | 0.578 |
| $a_{XM}^{bct}$ [nm] | 0.4121 | 0.4228 | 0.422 | 0.428 |
| $b_{XM}^{bct}$ [nm] | 0.2913 | 0.2788 | 0.270 | 0.271 |
| $c_{XM}^{bct}$ [nm] | 0.4093 | 0.4200 | 0.420 | 0.422 |
| $\beta_{XM}$ [1°] | 90.0 | 90.3 | 92.7 | 95.3 |
| $\lambda_1$ | 0.707 | 0.677 | 0.654 | 0.663 |
| $\lambda_2$ | 0.993 | 1.020 | 1.017 | 1.024 |
| $\lambda_3$ | 1.41 | 1.452 | 1.450 | 1.487 |
| $b_1$ | 0.408 | 0.430 | 0.422 | 0.431 |
| $b_2$ | -0.409 | -0.458 | -0.497 | -0.512 |
| $b_3$ | 0.409 | 0.454 | 0.458 | 0.437 |
| $m_1$ | 0.816 | 0.819 | 0.810 | 0.813 |
| $m_2$ | 0.408 | 0.401 | 0.376 | 0.335 |
| $m_3$ | -0.408 | -0.410 | -0.450 | -0.477 |